\documentclass[sigconf,authorversion]{acmart}
\AtBeginDocument{%
  \providecommand\BibTeX{{%
    \normalfont B\kern-0.5em{\scshape i\kern-0.25em b}\kern-0.8em\TeX}}}

\copyrightyear{2023}
\acmYear{2023}
\setcopyright{acmlicensed}\acmConference[CIKM '23]{Proceedings of the 32nd ACM International Conference on Information and Knowledge Management}{October 21--25, 2023}{Birmingham, United Kingdom}
\acmBooktitle{Proceedings of the 32nd ACM International Conference on Information and Knowledge Management (CIKM '23), October 21--25, 2023, Birmingham, United Kingdom}
\acmPrice{15.00}
\acmDOI{10.1145/3583780.3615187}
\acmISBN{979-8-4007-0124-5/23/10}

\usepackage{acronym} 
\usepackage{amsmath}
\usepackage{balance}
\usepackage[shortcuts]{extdash}
\usepackage{breqn}

\begin{document}

\acrodef{WTR}[WTR]{web table retrieval}
\acrodef{IR}[IR]{information retrieval}
\acrodef{LLM}[LLM]{Large Language Model}
\acrodefplural{LLM}[LLMs]{Large Language Models}
\acrodef{sDCG}[sDCG]{Session-based DCG}
\acrodef{idf}[idf]{inverse document frequency}
\acrodef{CSM}[CSM]{Complex Searcher Model}
\acrodef{SERP}[SERP]{search engine result page}
\title{Simulating Users in Interactive Web Table Retrieval}


 \author{Björn Engelmann}
 \affiliation{%
   \institution{TH Köln - University of Applied Sciences}
   \country{Germany}}
 \email{bjoern.engelmann@th-koeln.de}

 \author{Timo Breuer}
 \affiliation{%
   \institution{TH Köln - University of Applied Sciences}
   \country{Germany}}
 \email{timo.breuer@th-koeln.de}

 \author{Philipp Schaer}
 \affiliation{%
   \institution{TH Köln - University of Applied Sciences}
   \country{Germany}}
 \email{philipp.schaer@th-koeln.de}

\renewcommand{\shortauthors}{Björn Engelmann, Timo Breuer, \& Philipp Schaer}
\begin{abstract}

 Considering the multimodal signals of search items is beneficial for retrieval effectiveness. Especially in \ac{WTR} experiments, accounting for multimodal properties of tables boosts effectiveness. However, it still remains an open question how the single modalities affect user experience in particular. Previous work analyzed \ac{WTR} performance in ad-hoc retrieval benchmarks, which neglects interactive search behavior and limits the conclusion about the implications for real-world user environments.
 
 To this end, this work presents an in-depth evaluation of simulated interactive \ac{WTR} search sessions as a more cost-efficient and reproducible alternative to real user studies. As a first of its kind, we introduce interactive query reformulation strategies based on Doc2Query, incorporating cognitive states of simulated user knowledge. Our evaluations include two perspectives on user effectiveness by considering different cost paradigms, namely query-wise and time-oriented measures of effort. Our multi-perspective evaluation scheme reveals new insights about query strategies, the impact of modalities, and different user types in simulated \ac{WTR} search sessions.

\end{abstract}

\ccsdesc[500]{Information systems~Users and interactive retrieval}
\ccsdesc[500]{Information systems~Query reformulation}
\ccsdesc[300]{Human-centered computing}

\keywords{User Simulation, Interactive Table Retrieval, Multimodality, Query Generation}



\maketitle

\section{Introduction}
Web table retrieval is defined as finding relevant tables from a corpus for a given information need (typically expressed as a query).
This task is closely related to the classical retrieval of documents or passages.
For tables extracted from websites, however, the modalities, i.e., the contextual information, are of great importance.
This can be, e.g., the \textit{page title} from which the table originates.
While user simulation is an established practice in classical retrieval \cite{10.1145/2808194.2809470, 10.1145/3121050.3121070, 10.1145/3527546.3527559}, there is no research on web table retrieval so far.
However, this is particularly interesting since the impact of table modalities on users needs to be clarified.
In this paper, we combine insights from user simulation and table retrieval on the one hand and present new approaches to make interactive retrieval dynamic on the other hand. This dynamic component is based on the growing knowledge state of a user, where new queries depend on examined documents.




In sum, our contributions are as follows. We present a \textbf{1) query generation approach} based on Doc2Query for interactive user simulation. This is its first use for this kind of user simulation. Furthermore, we provide \textbf{2) an in-depth investigation of modality effects for table retrieval}. We analyze the information gain over an interactive course by modeling user behavior in various fashions. 
In our evaluation, we \textbf{3) contrast two cost paradigms}. The information gain is evaluated on the one hand --- classically throughout successive queries --- and, on the other hand, depending on the effort invested.
Thus, we gain realistic insights into our user sessions and a comprehensive picture of the simulation. This work is also the first study of its kind to run \textbf{4) an extensive simulation on an entire test collection}, including 60 topics. We also \textbf{5) make all resources used publicly available}.\footnote{\url{https://github.com/irgroup/simiir-wtr}} These include query variants of Doc2Query and by GPT-3.5 for every topic in the dataset, and the source code used.

\section{Related Work}
\label{sec:related_work}

Many approaches in \ac{WTR} have been presented, from classical retrieval methods based on term matching \cite{yanContentBasedTableRetrieval2017}, mapping tables or queries into semantic spaces \cite{DBLP:conf/sigir/0006B19}, to modern techniques based on large language models \cite{trabelsiStruBERTStructureawareBERT2022}.
Furthermore, it was investigated how the structure of the table and the context in which it appears can be exploited to improve the document representation for retrieval \cite{DBLP:conf/ictir/WangLWZYS15, DBLP:conf/sigir/ShragaRFC20}. 
Different datasets were used to compare these approaches. The Wiki Tables Test Corpus consists of 1.6M tables extracted from Wikipedia, for which 60 queries are available, along with relevance assessments \cite{zhangAdHocTable2018}. 
An extension is presented with the Web Table Retrieval Test collection \cite{DBLP:conf/sigir/ChenZ021}. Here, tables, with their context, are added from the Common Crawl dataset.
Tables are thus present in different modalities, such as \textit{page titles}, \textit{entities}, \textit{text before/after}, and the table itself.  Modality relevance assessments were created by crowdsourcing. 

Recently, Zobel~\cite{zobel2022} highlighted the ``gap'' between system\-/oriented measures and real user effectiveness. System-oriented measures are proxies that do not necessarily reflect the actual user effectiveness in real search situations. User variance is at least as important as system variance in retrieval evaluations~\cite{DBLP:conf/sigir/BaileyMST15}. For this reason, user simulations provide a promising solution to draw better conclusions about real-world implications in a cost-efficient way \cite{DBLP:journals/sigir/BalogMTZ21}. 

There exist several simulation framework modeling user behavior that generally agree on a core sequence of user interactions shared by all models~\cite{DBLP:conf/iiix/ThomasMBS14,DBLP:conf/sigir/MaxwellA16,DBLP:journals/ir/PaakkonenKKAMJ17,DBLP:conf/ictir/CarteretteBZ15,DBLP:conf/ictir/ZhangLZ17}. The user interaction sequence includes the \textit{1) query formulation} induced by the (topical) information need, \textit{2) scanning of the retrieved list} that is usually done by screening snippet texts, \textit{3) selecting and clicking} appealing items, \textit{4) reading documents and judging about their relevance}, and finally, \textit{5) inspecting other items}  in the result list and deciding about query reformulations or \textit{6) abandoning the search session} entirely.

There exist different ways to simulate queries if no real user logs are available~\cite{DBLP:conf/sigir/AzzopardiRB07,gunther2021assessing,DBLP:conf/cikm/BaskayaKJ13,DBLP:conf/ecir/PenhaCH22}. For instance, by generating queries from topic texts with principled rules~\cite{DBLP:conf/cikm/BaskayaKJ13}, making use of language models based on relevant documents to simulate queries for known-item search~\cite{DBLP:conf/jcdl/JordanWG06,DBLP:conf/sigir/AzzopardiRB07}, using query suggestions by Google~\cite{gunther2021assessing}, or fine-tuning \acp{LLM} with the help of topic texts and keyword queries as targets to generate new queries at inference time~\cite{DBLP:conf/ecir/PenhaCH22}. There are different ways to model browsing the snippets and making click decisions, e.g., by using editorial relevance labels to simulate click decisions based on probabilistic modeling~\cite{DBLP:conf/wsdm/HofmannSWR13} or parameterizing click models with the help of click logs~\cite{DBLP:series/synthesis/2015Chuklin}. Likewise, it is also possible to make click decisions based on language modeling approaches~\cite{maxwellAgentsSimulatedUsers2016}. In a similar way, relevance decisions about the entire document can be based on probabilistic modeling with editorial relevance labels or with language models. In this regard, it is also possible to incorporate reading time~\cite{DBLP:conf/sigir/SmuckerC12} and simulate different types of stopping behavior~\cite{DBLP:conf/ecir/MaxwellA18}. 

\section{Methodology}
Our approach focuses on user simulation and a feedback procedure. Since our analysis is concerned with tracking information gained over a sequential interactive course, we use a classical term matching-based approach, BM25, instead of neural methods that would maximize the effectiveness of ad-hoc retrieval. BM25 has already been used in various works for table retrieval and provides a well-established foundation.
We use the WTR dataset \cite{DBLP:conf/sigir/ChenZ021} and index across all modalities in our work. Both indexing and multi-field retrieval are implemented using PyTerrier \cite{DBLP:conf/cikm/MacdonaldTMO21}. For retrieval, we follow the methodology of \cite{DBLP:conf/sigir/ChenZ021}. Regarding the user simulation, we make use of the simulation toolkit \texttt{SIMIIR 2.0}~\cite{DBLP:conf/sigir/MaxwellA16,DBLP:conf/cikm/Zerhoudi0PBSHG22}.

We ground our user simulations on an adaption of the \ac{CSM}~\cite{DBLP:conf/sigir/MaxwellA16} that follows the interaction sequences outlined in the previous Section~\ref{sec:related_work}. \ac{CSM} implicitly assumes that the simulated users scan result lists and make click decisions based on the attractiveness of the search items' titles and snippets before judging about their relevance. 
In our experiments, we replace snippets with the different modalities that represent the table in the result list. If not specified otherwise, we assume that the users browse result lists with ten tables per \ac{SERP}, click on a table if its modality is relevant, use the \textit{page title} as the default modality, and the single table only adds up to the total information gain if it was not seen before. Most importantly, we put a special focus on the analysis of different modalities and how earlier seen tables affect generating query reformulations. 

\subsection{Query simulations}
\label{subsec:query_simulations}

As the \ac{WTR} dataset by Chen et al. does not provide any topic-related information besides the keyword-based queries, it is challenging to simulate queries without additional context information about the topics. Thus, we implement a query generation strategy that makes use of instruction-tuned LLMs and principled prompting.
More specifically, we prompt the \ac{LLM} to generate keyword queries with topic-adapted instructions. In our experiments, we query OpenAI's API and parse the outputs of \textbf{GPT-3.5} (more specifically, \texttt{gpt-3.5-turbo-0301}~\cite{openaiModelIndexResearchers2023}) based on the adaption of the following prompting template: \texttt{Please generate 100 keyword queries about <query>.}, where \texttt{<query>} is replaced with the query string of the particular topic. In total, we generate companion datasets with 100 query variants for each of the 60 topics that are publicly shared with the community for follow-up studies. In our simulation experiments, we treat the query sequence made by the \ac{LLM} as query reformulations. Following this approach, we intend to generate queries that are topically related but do not consider the context of earlier seen search results.

\textbf{Doc2Query}
Query generation with GPT-3.5 has the disadvantage that a fixed number of queries must be generated in advance, which do not refer to the domain-specific language of the data set and are independent of the search results.
For this reason, we present an approach that generates new queries based on a list of retrieval results (similar to pseudo relevance feedback). 
This approach is particularly suitable for user simulations in interactive information retrieval since each search iteration integrates the user's newly acquired knowledge.
We model this growing knowledge using keywords generated from seen tables.
We generate these keywords using the Doc2Query approach~\cite{DBLP:journals/corr/abs-1904-08375}.
Here, queries are predicted for a record (tables in our case) that are possible questions answered by the document.
The knowledge state $KS_i$ of a user after the $i$-th iteration (i.e., $i$-th query) is defined by the union of the terms that were generated from the seen tables using Doc2Query (\autoref{eq:knowledge_state}).

\begin{equation}
    \label{eq:knowledge_state}
    KS_i = \bigcup_{j \in \{1,\dots,i\}} \phi(\theta(Q_j)).
\end{equation}
\begin{equation}
    \label{eq:keywords}
    \phi(D) = \bigcup_{d \in D} \{t \in d2q(d) \,|\, idf(t) < 0.5 \land t \notin S \}.
\end{equation}

$\theta(Q)$ is the set of documents returned by a retrieval system for a given query $Q$.
We filter all stop words and terms whose \ac{idf} is less than 0.5.
The function $d2q(d)$ returns a set of terms retrieved from document $d$ using the Doc2Query approach.
These query terms are then used for subsequent iterations by adding a term from the knowledge state to the initial query $Q_0$. In this way, query variations are created that differ only in one term, are dependent on the search results, and represent the simulated user's knowledge.

\textbf{Feedback based Doc2Query}
Since the query simulation strategy is based only on tables without relevance judgments --- but our simulated user makes relevance judgments in each iteration --- we present a further approach that integrates these signals.
To account for user feedback in the form of relevance judgments, we define another knowledge state $KS_{i}^{rel}$, based only on terms obtained from documents that the simulated user has marked as relevant.

\begin{equation}
\label{eq:query_exp}
  Q_{i+1} = \left\{\begin{array}{ll}
        Q_0 \cup t^{rel} : t^{rel} \in KS_{i}^{rel} & KS_{i}^{rel} \neq \emptyset  \\ 
        Q_0 \cup t \; \; \; \; : t  \; \; \; \: \in KS_{i}  & else.
        
        \end{array} \right .
\end{equation}

This way, terms from relevant documents are used for the query variation.
If no new terms from relevant documents are available, the knowledge state $KS_i$ is used instead (\autoref{eq:query_exp}).

\subsection{Evaluation}

We conduct a multi-perspective cost analysis, including the \textit{query-} and \textit{time-wise} evaluation of the information gain. As proposed in earlier work, we evaluate the simulated sessions by the \ac{sDCG}~\cite{DBLP:conf/ecir/JarvelinPDN08}, which discounts the cumulative gain document- and also query-wise. \ac{sDCG} exclusively considers query formulations as \textit{costs} and thus models the stopping decision by a log-harmonic probability distribution over queries and documents.

We stress that users' search behavior typically covers more interactions that could result in \textit{costs}, e.g., inspecting snippets, reading full texts, and making judgments about relevance. To this end, we complement the \ac{sDCG} evaluations by considering all simulated actions in the simulated session logs to account for a more comprehensive perspective on the \textit{``effort vs. effect''} ratio. In our evaluations, we determine the effort by the passed time units (measured in seconds using the default configurations of \texttt{SIMIIR}) and compare them against the effect, i.e., information gain, based on the cumulated relevance scores of unseen documents.

\section{Experimental results}
\label{sec:experimental_results}

\begin{figure*}[!ht]
\centering
\includegraphics[width=.22\textwidth]{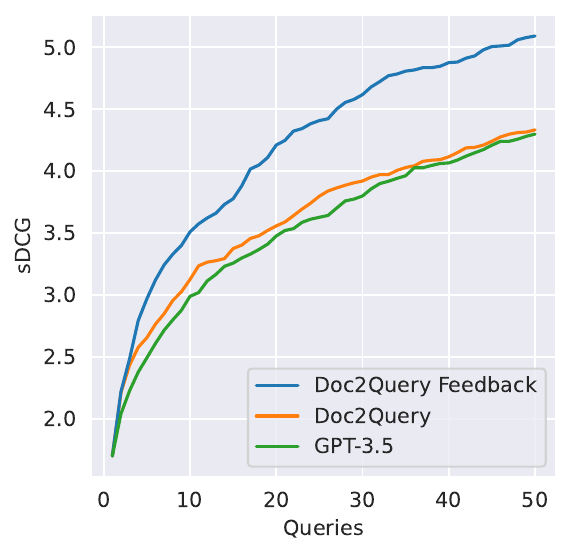}
\includegraphics[width=.22\textwidth]{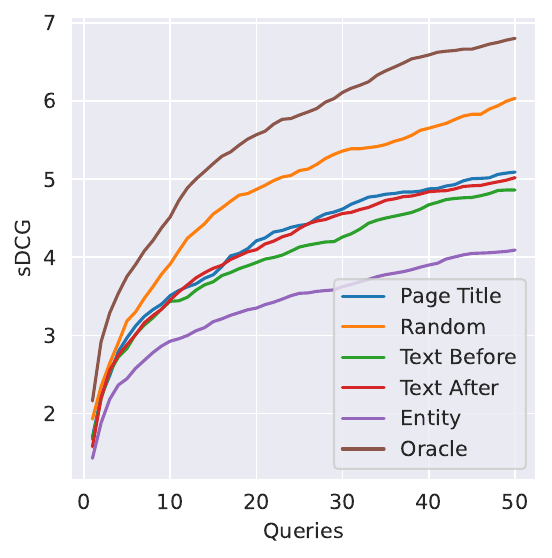}
\includegraphics[width=.22\textwidth]{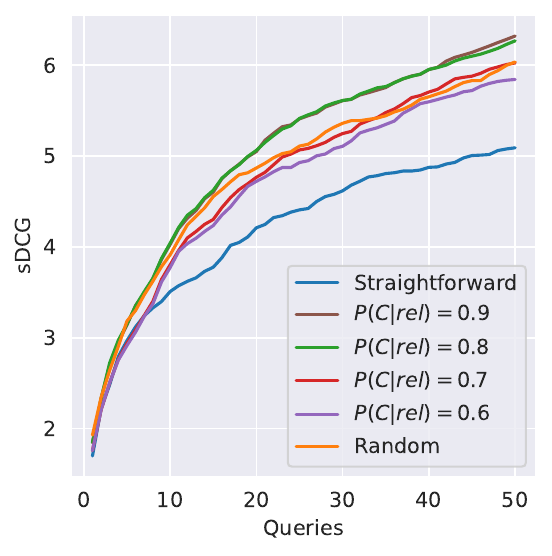}
\includegraphics[width=.22\textwidth]{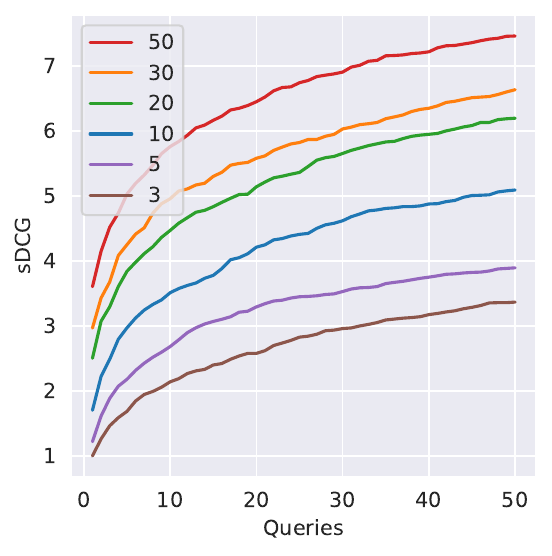}
\includegraphics[width=.22\textwidth]{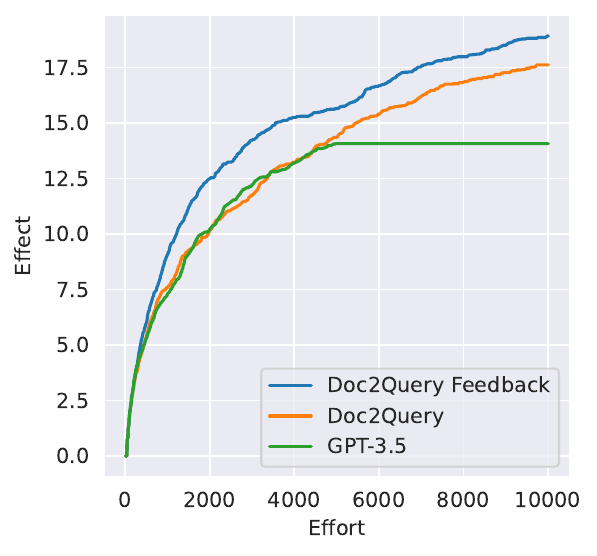}
\includegraphics[width=.22\textwidth]{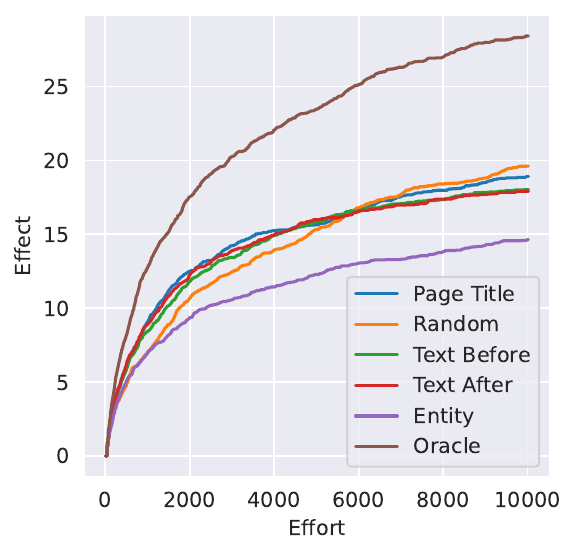}
\includegraphics[width=.22\textwidth]{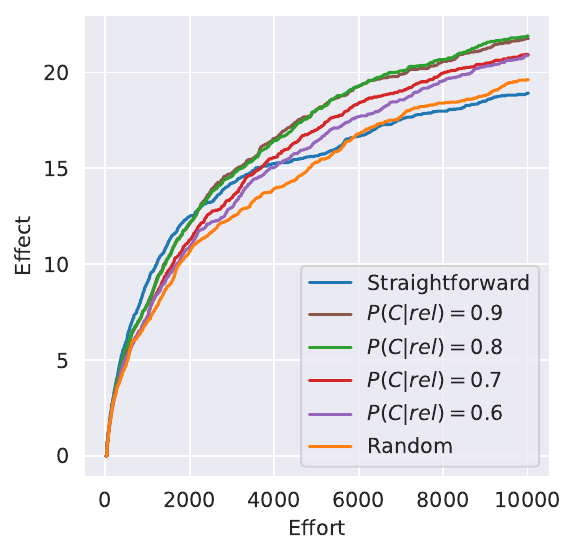}
\includegraphics[width=.22\textwidth]{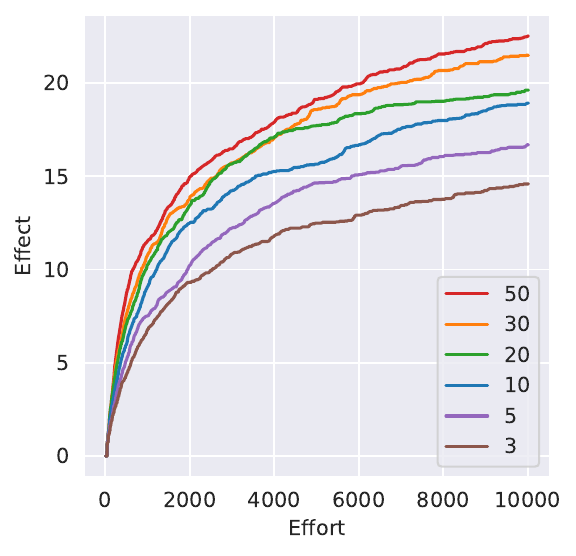}
\caption{Simulated web table retrieval sessions. Column-wise, the plots show the outcomes of four different evaluation levels. Row-wise, the plots can be compared by two evaluation paradigms. The blue lines represent the default parameters for all plots.}
\label{fig:experimental_results}
\end{figure*}

The following section describes the experimental results covering the comparison of querying strategies (cf.~\ref{subsec:query_strategies}), selection strategies based on different types of modalities (cf.~\ref{subsec:modalities}), and the general browsing behavior (cf.~\ref{subsec:browsing_strategies}). All of the results are visualized in Figure~\ref{fig:experimental_results}. The upper row contains plots based on a query-wise evaluation and the \ac{sDCG} measure. In contrast, the bottom row contains plots based on a more detailed resolution of effort and the resulting effects, i.e., information gain. Column-wise, the plots can be categorized into the four evaluation levels concerned with query strategy (first column), the modalities (second column), the click probabilities (third column), and browsing depths (fourth column). With this type of representation, the effects of the different evaluation levels can be compared horizontally, and the two cost paradigms can be compared vertically. 

\subsection{Comparison of Query Strategies}
\label{subsec:query_strategies}

The first column of plots in Figure~\ref{fig:experimental_results} compares simulated sessions with different types of query simulation strategies as introduced in Subsection~\ref{subsec:query_simulations}. Both plots show the increase in the average information gain over 60 topics with an increasing number of queries per topic. As can be seen, the feedback-based Doc2Query approach yields more effective sessions. By including key terms of relevant tables in later query reformulations, the simulated users pick up the terminology and better specify their information needs. In comparison, the GPT-based queries are less effective, which can be explained by the more generic way the queries were generated. Likewise, the Doc2Query-based query reformulations that consider all of the earlier retrieved tables for the reformulation result in lower effectiveness as they lack relevance feedback. There are no substantial differences between GPT and Doc2Query when no feedback is used. In conclusion, the instruction-tuned and task-agnostic GPT model achieves similar effectiveness as the task-specific Doc2Query.

By comparing both evaluation methods based on \ac{sDCG} and the time-wise information gain, we see similar results. However, our experiments show the limitations of using a predefined number of query reformulations that can be seen by the stagnating information gain for the GPT-based queries in the bottom plot. As the simulator runs out of queries, the search session ends, and there is no gain of information. In contrast, the Doc2Query-based generation method can be used for simulations with an arbitrary number of query reformulations. By these results, we conclude that the feedback-based Doc2Query method is suitable for reasonable user simulations, and we use it in all subsequent evaluations.

\subsection{Comparison of Modalities}
\label{subsec:modalities}

Since the feedback-based Doc2Query strategy was shown to be the most effective, the evaluations of the modality effects for this query expansion method follow in the second column of \autoref{fig:experimental_results}.
For this purpose, the tables' four modalities (\textit{page title}, \textit{before}, \textit{text after}, \textit{entity}), an Oracle, and a user with random behavior are examined.
Using the relevance scores of all modalities, we simulate users who only decide to click on a snippet based on its relevance and thus examine the complete table.
Furthermore, we add a user as an upper bound, the oracle, who already knows when viewing the snippet whether the table is relevant and only then clicks on the snippet. 
Additionally, there is a random behavior user who clicks on each snippet with a probability of 50\% ($P(C) = 0.5$).

Under both cost paradigms, the \textit{entity} modality performs worst as a relevance proxy. Furthermore, compared to the Oracle, a single modality does not serve as a good proxy for relevance. We suggest that relevance is a multidimensional concept and should not be represented by a single modality. The random behavior performs better than the modalities considering just the query-wise costs.  However, compared to the time-wise perspective, this behavior incurs higher costs and thus no longer performs better. In particular, at the beginning of the simulation, the random behavior performs worse than a user who uses the relevance signal of the \textit{page title}. We explain this effect by the fact that relevance proxies prevent the user from clicking on non-relevant tables at the beginning. On average, the user with random behavior clicks on half of the snippets in the SERPs and thus produces higher costs. In the course of the simulation, this becomes an advantage since he also finds relevant tables that a negative relevance signal of the snippets would hide. Our random user has a higher propensity to explore, while other users focus more on exploitation. 

\subsection{Comparison of Browsing Strategies}
\label{subsec:browsing_strategies}

The previous results indicate that random search behavior yields considerably effective outcomes, letting us conclude that a more in-depth evaluation concerning the \textit{exploitation/explorations} tradeoffs is required. To this end, the third column of Figure~\ref{fig:experimental_results} compares browsing strategies based on different click probabilities. We define the click probabilities by the snippet's relevance, as the random and straightforward click decisions are too simplistic and only reflect either fully exploratory- or exploitation-focused strategies.

The straightforward selection behavior implied that a table is clicked if its modality is relevant. At this evaluation level, we lower the click probability to simulate less strict exploitation-focused browsing strategies by combinations of $P(C | \mathrm{rel}) \in \{0.6, 0.7, 0.8, 0.9\}$ with $P(C | \neg \mathrm{rel}) = 0.3$, where $P(C | \mathrm{rel})$ and $P(C | \neg \mathrm{rel})$ denote the click probabilities for relevant or non-relevant modalities, respectively. For better comparability, we include the random ($P(C | \mathrm{rel}) = 0.5$, $P(C | \neg \mathrm{rel}) = 0.5$) and straightforward ($P(C | \mathrm{rel})=1.0$, $P(C | \neg \mathrm{rel})=0.0$) selection behaviors. 

As can be seen, the straightforward selection behavior is still the least effective strategy, especially when evaluated by \ac{sDCG} or as the search session tends to get longer. In contrast, most of the browsing strategies that relax the straightforward selection by slightly lower click probabilities perform best, which suggests that it is more effective to balance the strategies between exploitation- and exploration-focused browsing.

In conclusion, a single modality cannot comprise the entire notion of relevance that generally has to be understood as a multi-dimensional phenomenon. Random behavior is more effective, especially in the long run, which can be explained by \textit{serendipity} effects that occur by chance. However, it is more effective to apply a hybrid browsing strategy that emphasizes exploitation but which is also exploratory to some extent. 

We leave it as future work to analyze the combinations of modalities to simulate click decisions and to find the sweet spots for weighting exploitation and exploration. Finally, the fourth column of Figure~\ref{fig:experimental_results} shows the \ac{sDCG} and time-wise information gain for different browsing depths. As expected, users can either formulate more queries or browse more documents to increase their information gain, and consequently, the \ac{sDCG} curves with higher browsing depths lay above those with lower depths. However, these differences are less apparent as more interactions are considered as \textit{costs} as shown in the bottom plot. Naturally, browsing more documents also requires additional effort and time, which is not included in the \ac{sDCG}-based evaluations that solely model costs by the number of queries. Still, a higher browsing depth results in better overall effectiveness, but their advantages are less present early in the search sessions. 

\section{Conclusion}

Through this work, we introduce the first study of its kind that simulates interactive search sessions for web table retrieval. Furthermore, we introduced a query simulation method based on Doc2Query that can simulate an arbitrary number of queries by considering simulated relevance feedback. In this regard, we applied two different evaluation approaches based on query- and more comprehensive time-wise evaluations. Our results suggest that query-wise evaluations could be too simplistic as they only model queries as costs, and including other costs, such as scanning snippets and making click and relevance decisions, reveal a different picture of how different search strategies perform. Our modality-focused evaluations showed that there are differences between search effectiveness, and using a single modality as a snippet substitute is not recommended. Relevance is multi-dimensional and difficult to represent by a single modality. Future work should explore different combinations of modalities and analyze to which extent these could be used as proxies of the table's overall relevance in real user studies. 

\section*{ACKNOWLEDGEMENTS}
This work was supported by Klaus Tschira Stiftung (JoIE - 00.003.2020) and Deutsche Forschungsgemeinschaft (RESIRE - 509543643).

\bibliographystyle{ACM-Reference-Format}
\balance
\bibliography{cikm23-simulated-itr}

\end{document}